\documentclass{article}
\usepackage{}
\usepackage{amssymb}
\usepackage[top=12mm]{geometry}
\usepackage{amsmath,amsthm,amssymb}
\usepackage{fancyhdr}
\usepackage{graphicx}
\usepackage{hyperref}
\usepackage{lipsum}
\usepackage{subfig}
\usepackage{varwidth}
\usepackage{enumitem}
\usepackage{mdwlist}%
\usepackage[noadjust]{cite}%
\usepackage{array}
\usepackage{epstopdf}
\usepackage{tabularx}
\usepackage{caption}
\usepackage{color, soul}
\usepackage{ulem}
\usepackage{cancel}
\usepackage{textcomp}
\usepackage{float}
\title{ Direct observation of the CVD growth of monolayer MoS$_2$ using \textit{in-situ} optical spectroscopy}

\author{
       \small{C.B. L\'opez-Posadas$^1$, Yaxu Wei$^{1,2,3}$, Wanfu Shen$^{1,2,3}$, Daniel Kahr$^{1}$, Michael Hohage$^{1}$, Lidong Sun$^{1*}$}
        \\ \scriptsize \noindent\textit{1. Institute of Experimental Physics, Johannes Kepler University Linz, A-4040 Linz, Austria}
	\\ \scriptsize \noindent\textit {2. State Key Laboratory of Precision Measuring Technology and Instruments, Tianjin University, Weijin Road 92,}
	\\ \scriptsize \noindent\textit {Nankai District, CN-300072 Tianjin, China}
	\\ \scriptsize \noindent\textit {3. Nanchang Institute for Microtechnology of Tianjin University, Weijin Road 92, Nankai District, 300072 Tianjin, China}
	\\ \scriptsize\textit{* Corresponding author: lidong.sun@jku.at}}
\date{}
\begin{document}

	\maketitle
	\newcommand{\upcite}[1]{\textsuperscript{\textsuperscript{\cite{#1}}}}  
	\newcommand\hcancel[2][red]{\setbox0=\hbox{$#2$}%
	\rlap{\raisebox{.45\ht0}{\textcolor{#1}{\rule{\wd0}{1pt}}}}#2}
	\newcommand{\tabincell}[2]{\begin{tabular}{@{}#1@{}}#2\end{tabular}}
	\linespread{1.5}
    \renewcommand\thesection{\arabic{section}}

\section *{Abstract}	
Real-time monitoring is essential for understanding and eventually precise controlling of the growth of two dimensional transition-metal dichalcogenides (2D TMDCs). However, it is very challenging to carry out such kind of studies on chemical vapor deposition (CVD). Here, we report the first real time \textit{in-situ} study on the CVD growth of the 2D TMDCs. More specifically, CVD growth of molybdenum disulfide (MoS$_2$) monolayer on sapphire substrates has been monitored \textit{in-situ} using differential transmittance spectroscopy (DTS). The growth of the MoS$_2$ monolayer can be precisely followed by looking at the evolution of the characteristic optical features. Consequently, a strong correlation between the growth rate of MoS$_2$ monolayer and the temperature distribution in the CVD reactor has been revealed. Our result demonstrates the great potential of the real time \textit{in-situ} optical spectroscopy for the realization of the precisely controlled growth of 2D semiconductor materials.
\vspace{10pt} 

\section{Introduction}
The two-dimensional transition metal dichalcogenides (2D TMDCs) have drawn wide attention because of their fascinating physical and chemical properties~\cite{Novoselov2005, Mak2010, Wang2012, Butler2013, Xu2014, Manzeli2017}. Given that the potential advantages of 2D TMDCs as the active materials for various devices have been established, the primary requirement is cost-efficient, reliable, and high throughput synthesis of 2D TMDCs with processes compatible with the current semiconductor technology. Various approaches for synthesizing large-area 2D TMDCs have been reported, including mechanical exfoliations, sulphurization of metal thin films, mass transport, molecular beam epitaxy (MBE) and chemical vapor deposition (CVD)~\cite{Manzeli2017, Lin2016}. In particular, CVD is considered as an attractive and very promising approach for large-scale synthesis of 2D TMDCs~\cite{Lee2012, Lee2013, Wang2014, Ji2015, Dumcenco2015, Kang2015, Yu2017}. As a result, numerous empirical attempts have been taken to optimize the CVD process by choosing appropriate precursors, supporting substrate, carrier gases, flow rates, synthesis temperature, \emph{etc}. Although these studies have created helpful growth recipes and inspired many theoretical discussions about the growth mechanisms, they were still unable to provide a clear path to the fully understanding of the growth mechanisms~\cite{Wang2014, Cheng2015, Chen2015, Cain2016, Rajan2016, Kim2017}. Most recent efforts aiming at manipulating the nucleation and growth of 2D TMDCs have successfully led to the great improvement of the film quality and well orientated MoS$_2$ monolayer has even been fabricated in wafer scale~\cite{Dumcenco2015, Yu2017, Chen2015, Kim2017}. These results highlight the important role of surface kinetics in the CVD growth of the 2D TMDCs. To improve our understanding on the surface kinetics involved, \textit{in-situ} study performed during the CVD growth of 2D TMDCs is essential. However, according to the best of our knowledge, no study in this direction has been reported despite of its importance. Consequently, the growth mechanisms are still the subject of much speculation.

Because typical CVD used for the synthesis of 2D TMDCs involving gases with pressure in the range between atmosphere and hundredth of mbar, the characterization methods based on electron beam are not applicable and the \textit{in-situ} real time study becomes very challenging. On the other hand, each kind of 2D TMDCs possess specific optical properties which are characteristic of their chemical composition, crystalline structure and number of layers~\cite{Mak2010, Li2014}. Their optical properties are also sensitive to the extrinsic modifications, for instance, interaction with substrate~\cite{Jena2007}, gas molecule adsorption~\cite{Mouri2013, Tongay2013} as well as strain field~\cite{Conley2013}. These facts make the optical spectroscopy a sensitive probe to the surface kinetics involved during the growth of 2D TMDCs. Furthermore, the optical methods in visible range can be applied under various environments including vacuum, atmosphere and even high pressure condition. Consequently, the optical spectroscopy is the method of choice for \textit{in-situ} studies during the CVD growth. Monitoring the evolution of the optical properties of 2D TMDCs during CVD growth may provide a sensitive access to the surface processes and a characterization of morphology, crystalline structure of the films. By systematic study using \textit{in-situ} optical spectroscopy assisted with other \textit{ex-situ} characterizations, the details of the kinetics including adsorption, dissociation, reaction, nucleation and growth can be revealed.

Differential reflectance spectroscopy (DRS), which measures the normalized difference between the \textit{reflectance} of the bare surface and the one covered by thin films, posses enhanced sensitivity to the surface modification and ultrathin film growth~\cite{Forker2009,Bussetti2014}. The technique has been successfully applied to reveal the optical properties of 2D TMDCs~\cite{Mak2010, Li2014} and, most recently, also to monitor the molecular beam epitaxy of MoSe$_2$ monolayer on sapphire substrate~\cite{Wei2017}. In the current work, we have applied an analogous technology, namely, differential transmittance spectroscopy (DTS), to realize \textit{in-situ} real time study of the CVD growth of MoS$_2$ monolayer on Al$_2$O$_3$(0001) surface. In this case, the normalized differences between the \textit{transmittance} through the bare substrate and the one after a given deposition time are resolved. By monitoring the DTS during growth, the evolution of the optical properties associated to the MoS$_2$ layer can be revealed spectroscopically. Actually, for the van der Waals epitaxy of MoS$_2$ on the transparent sapphire substrate, the DT spectrum can be directly associated to the absorption of the adlayer. Consequently, the growth can be monitored \textit{in-situ} in real time and the detailed information associated to the kinetics can be deduced from the DT spectra.

\section{Results and Discussions}

\subsection{Ex-situ characterization}

\begin{figure}[t!]
	\centering
	\includegraphics[width=0.9\textwidth]{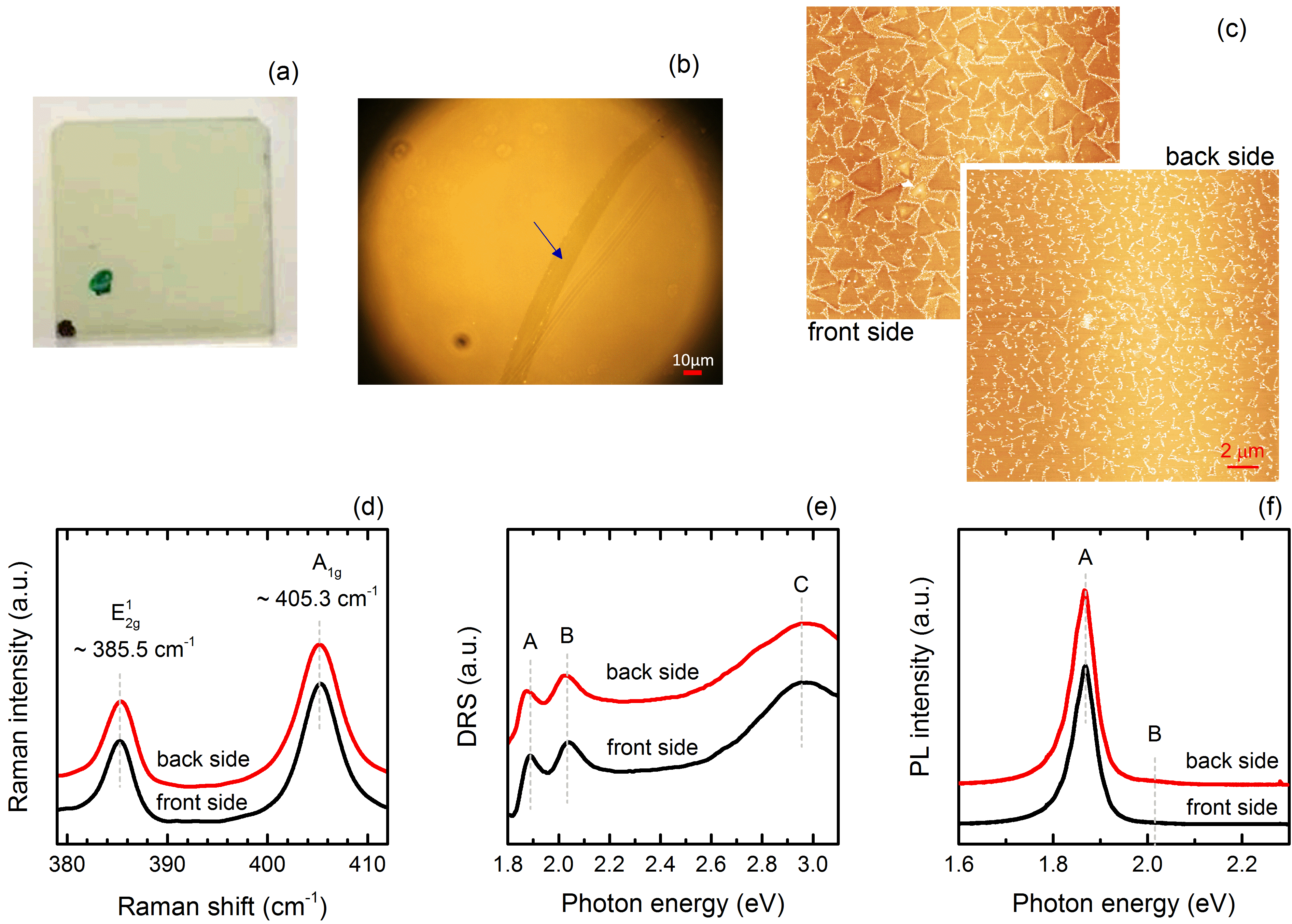}
	\caption{The results of the \textit{ex-situ} characterizations. (a) The photo representing the 10\,mm $\times$ 10\,mm substrate after the CVD growth. The black and green dots marked on the front and back sides of the substrate, respectively, are used to distinguish between the two surfaces. (b) Optical microscopy image taken from the surface at the front side of the same substrate. A scratch is indicated by the arrow. (c) AFM images recorded from the surfaces at front and back side of the substrate, respectively. (d), (e) and (f) The Raman, DR and PL spectra recorded on the surfaces at front and back side of the substrate, respectively. }
	\label{fig:Ex-situ}
\end{figure}

Following the CVD process described in the Experimental section, MoS$_2$ monolayer has been deposited, simultaneously, on not only the surface at the front side (face to the Ar flow direction) but also the one at the back side of the sapphire substrate. This conclusion is based on a thorough \textit{ex-situ} characterization after CVD growth using differential reflectance spectroscopy (DRS), Raman spectroscopy, photoluminescent spectroscopy (PL), optical microscopy (OM) and atomic force microscopy (AFM). Actually, from the first glance at the substrate after CVD (see Fig.\,1(a)), it can be recognized that the substrate shows a homogeneous light green colour which is characteristic for the sapphire substrate covered by MoS$_2$ monolayer. Indeed, OM images (Fig.\,1(b)) taken at different areas of the front and back sides of the substrate reveal that the surfaces at both sides are covered by homogeneous layers which can only be recognized at the appearance of defects. Fig.\,1(c) presents the AFM images measured at front and back sides of the sapphire substrate confirming both surfaces are covered by an almost completed monolayer of MoS$_2$. The Raman spectra recorded from each side of the substrate (Fig.\,1(d)) show the characteristic peaks at 385.5 and 405.3\,cm$^{-1}$, which are attributed to the in-plane (E$^1_{2g}$) and out-of-plane (A$_{1g}$) vibration mode of the 2H MoS$_2$ crystal. Most importantly, the interval between these two peaks is $\sim$ 20\,cm$^{-1}$, which is characteristic for the MoS$_2$ monolayer~\cite{Tonndorf2013}. The DR spectra measured at the surfaces on the front and back sides of the substrate are plotted in Fig.\,1(e). The peaks marked by A and B, which are located around 1.9 and 2\,eV, are due to the excitonic transitions occurring at the K and K$^{\prime}$ points of the Brillouin zone, respectively. The broad peak C around 2.95\,eV is attributed to the interband transitions transpiring near the critical point of $\Gamma$, where the valance and conduction bands are nested~\cite{Qiu2013}. The relaxation of the transitions A and B gives strong PL emission which are plotted in Fig.\,1(f). In order to check the homogeneity, all the measurements have been performed at several different spots over the substrate surfaces and the results are rather identical. Consequently, homogeneous MoS$_2$ monolayer has been synthesized on both front and back sides of the substrate. However, a closer inspection reveals some detailed difference between the morphology at front and back surfaces regarding the surface coverage and the grain size. The grains grow on the back surface have already coalesced forming a rather compact monolayer. In contrast, the grains on the front surface are still rather separated. Furthermore, it appears that grain size is relatively large on the front surface. These observations indicate a difference between the front and back side of the substrate regarding the growth kinetics. We attribute the observed distinction tentatively to the possible deviation of the effective deposition rate on each side of the substrate. Actually, the deposition rate could be enhanced on the back surface due to the interruption of the gas flow by the substrate which may introduce a local turbulence and in turn a higher deposition rate.

\subsection{In-situ real time measurement}

\begin{figure}
	\centering
	\includegraphics[width=0.9\textwidth]{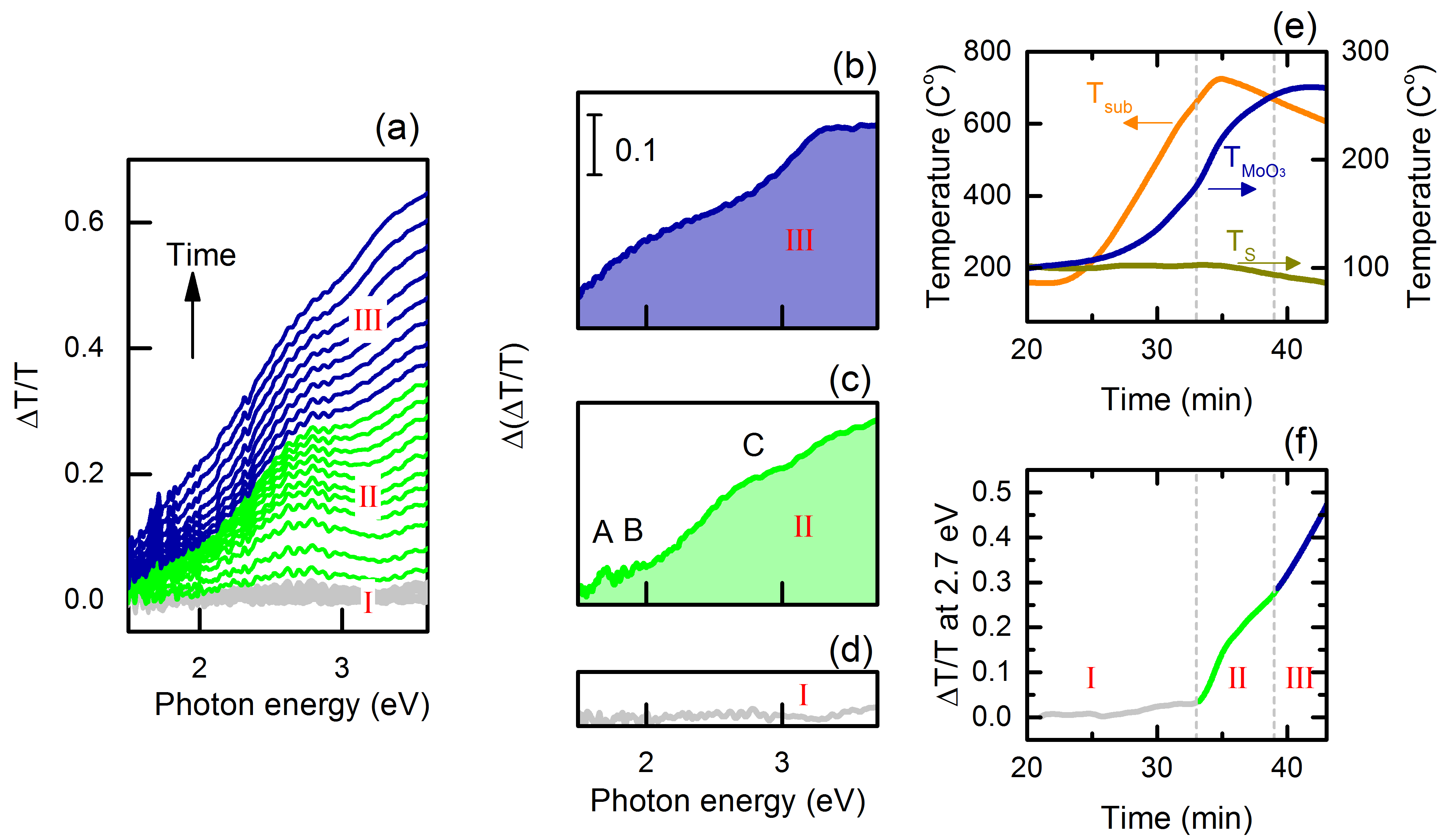}
	\caption{(a) DT spectra ($\Delta T/T$) recorded during the CVD growth. The time interval between the adjacent spectra is 30\,s. (b), (c) and (d) The incremental changes of the DT spectrum over the periods of third, second and first section of the growth, respectively. (e) The temperature profiles of the substrate ($T_{sub}$), MoO$_3$ source ($T_{MoO_3}$) and the sulfur source ($T_S$) recorded during growth. (f) The corresponding DT intensity at 2.7\,eV recorded during CVD growth.}
	\label{fig:DTS}
\end{figure}

Next, let's have a look at the evolution of the optical properties during the growth. Fig.\,2(a) shows the DT spectra recorded during the CVD process. Based on the evolution of the DTS, we divide our CVD process into three sections. For the first section (section I) starting from the preheating of the sulfur source and the furnace, it can be seen that the DT spectrum remains unchanged during the first 33\,min, although the temperatures measured at the positions of sulfur, MoO$_3$ and the substrate were raised. This observation shows that no thin film was deposited during this period. The second section (section II) starting from the time $t\,=\,33$\,min, transmittance around 2.7\,eV decreases (DT signal increases) with the time. The energy position of the DTS feature at 2.7\,eV coincides very nicely with the absorption peak C of the MoS$_2$ monolayer. The continuous decrease of the optical intensity at this energy indicates the increase of the optical absorption due to the growth of the MoS$_2$ layer on the surfaces of the sapphire substrate. In comparison with the DR spectra of the same sample measured \textit{ex-situ} at room temperature (see Fig.\,1(e)), the peak C is broad and shifted to the lower energy. Furthermore, the sharp absorption features A and B observed at room temperature are hardly visible. We attribute this deviation from the optical spectrum measured at room temperature to the high substrate temperature ($\sim$ 700\,$^{\circ}$C) during growth. Indeed, the study of the temperature dependence of the absorption of MoS$_2$ monolayer shows the same tendency~\cite{Yalon2017}. Therefore, the DT spectra recorded after more than 33\,min deposition show characteristic absorption of MoS$_2$ monolayer at elevated temperature and their amplitudes increase with the deposition time. However, during the third section (section III), which starts at about 39\,min, the DT signal at a higher energy of 3.2\,eV became dominating. Based on its energy position, this feature can be attributed to the absorption of the MoO$_3$~\cite{Itoh2001, Castro2017}. As a semiconductor, MoO$_3$ shows a bandgap around 3\,eV. Considering the high temperature of the substrate, we can exclude the possibility that the MoO$_3$ thin films were formed on the substrate surfaces. Actually, we attribute the observed DTS signal to the absorption of the MoO$_3$ thin films which were deposited on the inner surface of the quartz window at the end of the furnace tube. Since the quartz was maintained close to room temperature, it is reasonable to assume that a small amount of the MoO$_3$ molecules carried by the Ar flow from upstream may be deposited on the inner surface of this quartz window. The deposition rate of the MoO$_3$ on the inner surface of the quartz window should be proportional to the product of $p \cdot C_{MoO_3}(T_{MoO_3})$, in which, $C_{MoO_3}(T_{MoO_3})$ is the concentration of MoO$_3$ molecule above the MoO$_3$ source which depends exponentially on the source temperature $T_{MoO_3}$. The coefficient $p$ count on the transportation efficiency of MoO$_3$ molecules from the source to the inner surface of the quartz window. Although the coefficient $p$ has been minimized, it was, apparently, unignorable during this study.  Actually, the growth of the MoO$_3$ feature sets in already in the second section, where the DT signal raising at 2.7\,eV was accompanied by the increasing at 3.2\,eV.

The characteristic increments of the DT spectrum during each section can be recognized from the $\Delta(\Delta T / T)$ spectra plotted in the central column of Fig.\,2. Indeed, the increment of the DT signal over the large wavelength range is negligible during the section I (Fig.\,2(d)) indicating no growth occurring on the substrate surfaces. For the section II (Fig.\,2(c)), the increase of the DT signals can be clearly recognized at not only 2.7\,eV but also 1.85 and 1.72\,eV associating to the absorption peaks C, B and A of the growing MoS$_2$ monolayer, respectively. In comparison with the DR spectra measured \textit{ex-situ} at room temperature, a systematic red shift can be observed for all the three features verifying the influence of the substrate temperature. In addition, the absorption of the MoO$_3$ thin films deposited on the quartz window is also clearly visible at 3.2\,eV. Finally, during the section III (Fig.\,2(b)), the incremental change of the DTS is dominated by the increase in the energy range above 3.2\,eV. In contrast, the features associated to MoS$_2$ becomes invisible. The decrease of the transmittance (the increase of DTS) over the low energy range could be attributed to the increase of the scattering of the MoO$_3$ thin film deposited on the quartz window.

In order to understand the observed evolution of the DT spectroscopy, let's have a closer inspection at its correlation with the temperature at the positions of sulfur source, MoO$_3$ source and the substrate (see Fig.\,2(e) and (f)). In Fig.\,2(f), the DT signal at 2.7\,eV is selected to represent the growth of the MoS$_2$. By looking at Fig.\,2(e) and (f), it becomes clear that the growth of the DT signal at 2.7\,eV and thus the MoS$_2$ layer sets in when the temperatures of the substrate and MoO$_3$ exceed 650\,$^{\circ}$C and 170\,$^{\circ}$C, respectively. Based on the following two facts: (1) The sulfur source had been being maintained at temperatures above 100\,$^{\circ}$C for more than 10\,min till this moment. (2) The growth of MoS$_2$ monolayer at substrate temperature as low as 530\,$^{\circ}$C has been reported~\cite{Ji2015}. We thus attribute the observed onset of the MoS$_2$ growth to the evaporation of the MoO$_3$ reaching a recognizable rate. Afterwards, the DT signal at 2.7\,eV increases monotonically revealing the ongoing growth of the MoS$_2$ layer. Interestingly, although the temperature of the MoO$_3$ source was still increasing, the growth speed of the DT signal at 2.7\,eV dropped at around 35\,min. With a closer inspection of the temperature curves, the observed decrease of the growth rate at 2.7\,eV can be correlated to the declining of the substrate temperature and the sulfur source temperature. Actually, without going into the details of the kinetics of the reaction, the growth rate $r$ of the MoS$_2$ can be associated to several factors that depend strongly on the temperatures of sulfur source ($T_{S}$), MoO$_3$ source ($T_{MoO_3}$) and the substrate ($T_{sub}$), respectively, in the following way:
\begin{equation}
r \propto k(T_{sub}) \cdot C_{MoO_3}(T_{MoO_3}) \cdot C_{S}(T_{S}),
\label{eq:Reaction}
\end{equation}
where $k(T_{sub})$ represents the effective reaction rate coefficient between MoO$_3$ and sulfur leading to the formation of MoS$_2$. The $k(T_{sub})$ depends strongly on the $T_{sub}$ as indicated by the Arrhenius equation. Similar to $C_{MoO_3}(T_{MoO_3})$, $C_{S}(T_{S})$ is the concentration of the sulfur molecules over the sulfur source, which depends exponentially on $T_{S}$. This relation emphases the strong influence of the temperature on the growth rate of MoS$_2$. Indeed, the fast rise of the DT signal at 2.7\,eV at the initial stage of the section II can be correlated nicely to the increase of the temperatures at both MoO$_3$ source ($T_{MoO_3}$) and around the substrate ($T_{sub}$). The drop of the growth rate of the DT signal at 2.7\,eV observed at $\sim$35\,min coincides decently to the decrease of the temperature around the substrate ($T_{sub}$) and at the source of sulfur ($T_{S}$). The recovering of the growth rate of DT signal at 2.7\,eV within the section III can not be attributed to the absorption of MoS$_2$ anymore. Instead, it is induced by the overall decrease of transmittance through the quartz window due to the enhanced coating of the MoO$_3$ layer. Indeed, in contrast to the decrease of $T_{sub}$ and $T_S$, the MoO$_3$ source temperature $T_{MoO_3}$ keeps rising even at the beginning of the section III (see Fig.\,3(e)).

\section{Conclusion}

In conclusion, we have grown successfully the MoS$_2$ monolayers on both sides of the double polished c-plane sapphire substrates using CVD. The monolayers on both sides of the substrate distributed homogeneous over the substrate surfaces with a size of 10\,mm $\times$ 10\,mm. Most importantly, the evolution of the optical transmittance of the substrate has been monitored \textit{in-situ} in real time during the CVD growth using DTS. The formation of the MoS$_2$ monolayers is clearly visible from the development of DT spectrum. More specifically, the onset of the growth and the variation of the growth rate of MoS$_2$ monolayers can be determined. These detailed information about the growth deduced from the DTS can be correlated very nicely to the variation of the temperatures of the substrate, the MoO$_3$ and the sulfur sources. The current result emphasizes the importance of the \textit{in-situ} real time study and paves the way for deep understanding and eventually precise controlling of the growth of the 2D semiconducting materials.

\section{Experimental}

\begin{figure}
	\centering
	\includegraphics[width=0.9\textwidth]{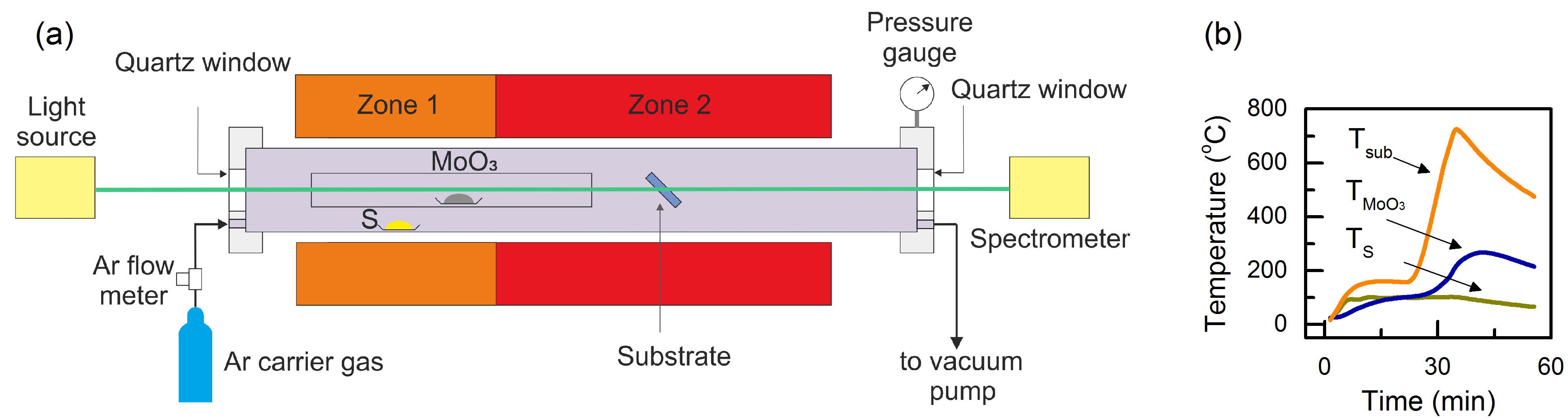}
	\caption{(a) The schematic for the CVD growth of monolayer MoS$_2$. (b) Temperature profiles for the substrate, MoO$_3$ and sulfur. }
	\label{setup}
\end{figure}

The setup of the CVD reactor including the optical spectrometer applied for the \textit{in-situ} measurement is sketched in Fig.\,3(a). High purity MoO$_3$ powder is used as the precursors for molybdenum, whereas the sulfer is supplied by vaporizing solid sulfur. Ar with a high purity was used as carrier gas and a flow rate of 150\,sccm was maintained during the whole process. The air in the reactor was evacuated before the inlet of the Ar gas and the pressure of Ar was maintained at 0.1\,torr till the end of the process. An horizontal tube furnace with a single heating zone and a heating belt were applied as heating elements for the substrate, MoO$_3$ and sulfur, respectively. To establish the temperature distribution required for the CVD, the substrates were positioned at the center of the furnace, while the alumina boats containing solid sulfur and MoO$_3$ were located 15\,cm and 3\,cm, respectively, away from the entrance of the furnace upstream of the Ar flow. From the configuration exhibited in Fig.\,3(a), it can be recognized that the sulfur is predominantly heated by the heating belt, whereas the MoO$_3$ powder was heated not only by the heating belt but also by the heater in the furnace in a radiative manner. In order to avoid the early mixing and reaction between MoO$_3$ and sulfur, the alumina boat of MoO$_3$ was enclosed in a quartz tube suspended coaxially with the outer quartz tube of the CVD reactor. The distance between the exit of the inner quartz tube and the substrate was maintained as 3\,cm. The evolutions of temperatures of the substrate, boats containing MoO$_3$ and sulfur during growth are plotted in Fig.3(b). For the growth reported in this paper, double side polished c-plane sapphire (Al$_2$O$_3$(0001)) was selected as the substrate. During deposition, the substrates were inclined with their surface normal 45 degree off the axis of the quartz tube of the furnace (see Fig.\,3(a)). At both ends of the furnace quartz tube, windows made of quartz were mounted to seal the reactor. The optical transmittance of the substrate was measured in real time using a spectroscopic ellipsometer from J.A. Woollam Co. More specifically, a beam of white light generated from a Xe lamp was guided through the substrate along the axis of the CVD reactor and detected by an spectrometer equipped with a CCD detector array. In order to enhance the sensitivity, the differential transmittance spectrum (DTS) at time $t$ was calculated using the following equation (Eq.~\ref{eq:DTS}):

\begin{equation}
\frac{\Delta T}{T} (t) = \frac{T_0-T_t}{T_0},
\label{eq:DTS}
\end{equation}
where $T_0$ and $T_t$ denote the transmittance spectra of the bare substrate before the CVD process and the one after a time $t$, respectively. The obtained DTS signal $\frac{\Delta T}{T} (t)$ represents the change of the optical transmittance relative to the bare substrate surfaces as a function of time $t$. Because the sapphire substrate is transparent in the wavelength range studied here, the DT spectra measured are directly associated with the absorption of the thin films deposited on the substrate. This fact makes the DTS a sensitive method for \textit{in-situ} monitoring the growth of MoS$_2$ monolayer in real time in the current study.

\section*{acknowledgements}
	C.B.-P and Y. W contribute equally to this work.The authors thank Assoc.Prof. Chunguang Hu and Prof. Yanning Li at Tianjin University for fruitful
	discussions.We acknowledge financial support for this work by the Austrian Science Fund (FWF) with project number: P25377-N20. C.B.L.P. thanks financial support from Consejo Nacional de ciencia y Tecnolog\'ia, though the ``Becas Mixtas'' program. Y.X.W. acknowledges the financial support from Eurasia Pacific Uninet and W.F.S. acknowledges the financial support of China Scholarship Council (CSC).

\scriptsize
\bibliographystyle{unsrt}
\bibliography{Bib}

\end{document}